\documentclass[12pt]{article}

\newenvironment{pf}{\noindent\textbf{Proof.}~}{\hfill{$\Box$}}

\usepackage{graphicx}
\usepackage{epsfig}
\usepackage{rotating}
\usepackage{amssymb}
\usepackage{amsmath,amsfonts,mathtools}
\usepackage[latin1]{inputenc}
\usepackage{verbatim}
\usepackage[tableposition=top,font=small,labelfont=bf,format=hang]{caption}
\usepackage{booktabs}
\usepackage{placeins}
\usepackage{needspace}
\usepackage[T1]{fontenc}
\usepackage{lmodern}
\usepackage{microtype}
\usepackage[hidelinks,pdfusetitle]{hyperref}

\numberwithin{equation}{section}

\newtheorem{defi}{Definition}
\newtheorem{lem}{Lemma}
\newtheorem{thm}{Theorem}
\newtheorem{prop}{Proposition}

\newtheorem{ex}{Example}

\newcommand{\F}{\mathbb{F}}

\newcommand{\Z}{\mathbb{Z}}

\DeclareMathOperator{\wt}{wt}

\newcommand{\one}{\mathbf1}

\title{A construction of optimal single-track Gray codes of length \texorpdfstring{$2p$}{2p}\thanks{The work of Minjia Shi was supported by the National Natural Science Foundation of China under Grants 12471490 and W2611002. ({\em Corresponding author: Minjia Shi})}}
\author{Minjia Shi\thanks{Key Laboratory of Intelligent Computing Signal Processing, Ministry of Education, School of Mathematical Sciences, Anhui University, Hefei 230601, China.  Email: \texttt{smjwcl.good@163.com}} \and Xinguang Dai\thanks{School of Mathematical Sciences, Anhui
		University, Hefei 230601, China. Email: \texttt{daixg88@163.com}}}
\date{}

\begin{document}
\maketitle
\begin{abstract}
A single-track Gray code is a cyclic Gray code in which each track
is a cyclic shift of the first. We solve the following conjecture suggested by Schwartz and Etzion in
1999: for every odd prime $p$, does
there exist a binary single-track Gray code of length $2p$ attaining
the maximum possible period $2^{2p}-4$? We construct such a code by ordering necklace representatives along
an Euler circuit in a directed multigraph.
\end{abstract}


\medskip
\noindent
{\bf Math Sc. Classification (2020):} 05C45, 05C70, 94A55.

\section{Introduction}\label{sec:introduction}

A binary Gray code lists distinct $n$-bit words so that consecutive
words differ in exactly one bit~\cite{Savage}. It is cyclic if the
last and first words also differ in one bit.
In angular quantization, only the bit that changes can be
ambiguous at the boundary between adjacent intervals~\cite{Gilbert1958}. A cyclic Gray code with
$P$ words is a \emph{single-track Gray code} (STGC) of period $P$ if,
when its words are written as rows of an array, each column is a cyclic
shift of the first~\cite{HPB}. The set of all cyclic shifts of a binary
word is called a binary \emph{necklace}~\cite{DD}. The defining property of STGCs offers profound practical advantages: by encoding all information onto a single physical track, they drastically reduce the hardware footprint and completely eliminate the need for multiple concentric tracks. This makes them exceptionally well-suited for miniaturized devices. STGCs have a rich history in position sensing and analog-to-digital conversion, with applications spanning from absolute rotary and linear encoders~\cite{Brandestini1995,EP,SE} to emerging spintronic devices~\cite{Yan2025}. For instance, Zhang et al.~\cite{Zhang} demonstrated a rotary encoder using eight read heads on a single track to resolve 128 angular positions per revolution. Extending this paradigm to modern hardware, Yan et al.~\cite{Yan2025} recently proposed a low-power 8-bit spintronic encoder that digitizes an input current via the angular displacement of a magnetic domain wall. A fundamental requirement across these diverse applications is that a longer period \(P\) allows for finer resolution (more positions or signal levels) without increasing the number of read heads (length \(n\)). To address such practical constraints, computational methods have been explored. For instance, 
Fakhri and Mallah Livani~\cite{FakhriLivani} used mixed-integer linear programming to search for STGCs of prescribed periods. Therefore, to fully satisfy this inherent engineering demand for maximum information density, the theoretical pursuit of \emph{optimal} binary STGCs---which attain the absolute largest possible period for a given length---becomes critically important.

The development of constructing optimal STGCs has been significantly advanced by the pioneering and seminal contributions of Etzion and his collaborators. By ingeniously ordering representatives of binary necklaces, Etzion and Paterson~\cite{EP} constructed optimal STGCs of lengths 9, 11, and 13. Furthermore, through the strategic application of self-dual necklaces, they provided a remarkable recursive construction yielding an infinite family of STGCs of length \(n=2^m\) and period \(2^n-2n\) for every \(m \ge 3\). This profound insight was later validated by Schwartz and Etzion~\cite{SE}, who proved that \(2^n-2n\) is indeed the maximum possible period for these lengths, firmly establishing the optimality of this infinite family. Etzion~\cite{E} later extended this approach to nonbinary alphabets
using nonbinary self-dual sequences. For $q\in\{3,5\}$ and every
$t\ge 1$, his construction gives a $q$-ary STGC of length $n=q^t$
that attains the full period $q^n$, further propelling the ongoing development of STGCs.

Beyond the established optimal constructions, the study of STGCs still leaves some fundamental open problems. Specifically, Schwartz and Etzion~\cite{SE} conjectured that full-order self-dual
necklaces of length $4p$ could yield a binary STGC of length $2p$
and period $2^{2p}-4$ for every odd prime $p$. This problem was also posed in~\cite[Problem~7.1]{Etzion2024}. We solve this conjecture as follows. For every odd prime $p$, the orbits
of size $4p$ under the complemented cycling register $T$ contain all
but four binary words of length $2p$~\cite[Theorem~6]{E}. We represent
these orbits by the arcs of a directed multigraph. At each nonzero
vertex, the permitted transitions form a $p$-dimensional cube. We use
Fink's matching theorem~\cite{Fink} on these cubes to join the circuits
determined by local matchings into a single Euler circuit. Along this
circuit we choose one representative from each orbit so that successive
representatives differ in one bit. A choice of alignment at the zero
vertex, followed by a relabeling of the coordinates, ensures that the
last representative also differs in one bit from the $T$-shift of the
first. Concatenating the $4p$ successive $T$-shifts of this list gives
a binary STGC of period $2^{2p}-4$.

Section~\ref{sec:preliminaries} recalls the period bound, the self-dual
necklace construction, and the matching theorems used in the proof.
Section~\ref{sec:orbits} classifies the orbits of the complemented
cycling register at length $2p$. Section~\ref{sec:transitions} defines
the permitted transitions and constructs the Euler circuit. In
Section~\ref{sec:construction} we prove optimality, give the
construction algorithm, and work out the length-six example.
Section~\ref{sec:conclusion} concludes the paper.
\section{Preliminaries}\label{sec:preliminaries}

We write $\F_2$ for the binary field. Addition of binary words is
coordinatewise over $\F_2$. For a binary word $u$, let $\wt(u)$ denote
its Hamming weight and let $\bar u=u+\one$ denote its complement, where
$\one$ is the all-one word of the appropriate length. The Hamming
distance between $u$ and $v$ is $d_H(u,v)=\wt(u+v)$. We identify a
pair of words with their concatenation, so
\[
 d_H\bigl((u,v),(u',v')\bigr)=d_H(u,u')+d_H(v,v').
\]
A \emph{Hamming isometry} is a bijection that preserves Hamming
distance. Coordinate permutations and complementation in any fixed
set of coordinates are Hamming isometries.

\subsection{Single-track Gray codes}
We recall the definition of a binary STGC and the bound on its period.
\begin{defi}\normalfont
A \emph{binary single-track Gray code} of length $n$ and period $P$ is a
cyclic list $C_0,C_1,\ldots,C_{P-1}$ of distinct words in $\F_2^n$ such
that $d_H(C_i,C_{i+1})=1$ and, for every $j$, there is an integer $h_j$
satisfying
\begin{equation}\label{eq:single-track}
 (C_i)_j=(C_{i+h_j})_0\ (0\le i<P,\ 0\le j<n).
\end{equation}
Row indices are taken modulo $P$. The integers $h_j$, with $h_0=0$, are
the \emph{head positions}, and the zeroth coordinate sequence is the
\emph{generating track}. A code is \emph{optimal} if its period is largest
among all single-track Gray codes of the same length.
\end{defi}

\begin{prop}[{\cite[Lemma~2]{HPB}; \cite[Lemma~1]{SE}}]\label{prop:bound}
The period $P$ of a binary single-track Gray code of length $n$ satisfies
$2n\mid P$ and $P\le 2^n$. In particular, if $n=2p$ for an odd prime
$p$, then $P\le 2^{2p}-4$.
\end{prop}

\subsection{Orbits and the complemented cycling register}
\label{subsec:register}

Let $G$ be a group of permutations of a finite set $\Omega$. The
\emph{orbit} of $x\in\Omega$ under $G$ is $Gx=\{gx:g\in G\}$, and
$\Omega/G$ denotes the set of these orbits. We write
$\langle U_1,\ldots,U_k\rangle$ for the group generated by the
permutations $U_1,\ldots,U_k$. In particular, the orbit under a
permutation $U$ is
\[
 \operatorname{Orb}_U(x)=\langle U\rangle x=\{U^jx:j\in\Z\}.
\]

If $f:\Omega\longrightarrow\Omega'$ is a bijection and
$V=fUf^{-1}$, then $V^jf(x)=f(U^jx)$ for every integer $j$.
Thus $f$ gives a bijection between the $U$-orbits and the $V$-orbits,
preserving their sizes. We also use the following elementary property:
if $U$ and $V$ commute and have coprime orders $a$ and $b$, then $UV$
has order $ab$ and $\langle UV\rangle=\langle U,V\rangle$.

The complemented cycling register on $\F_2^n$ is the map
\begin{equation}\label{eq:register}
 T(x_0,x_1,\ldots,x_{n-1})=(x_1,x_2,\ldots,x_{n-1},1-x_0).
\end{equation}
It is a Hamming isometry, $T^n x=\bar x$, and $T^{2n}x=x$.
A $T$-orbit of size $2n$ is called a \emph{full orbit}.
A binary necklace is \emph{self-dual} if it is invariant
under bitwise complementation~\cite{Etzion1987}. The following
lemma is the self-dual word construction of Etzion and Paterson with
shift one, expressed in terms of $T$. A seed $X$ corresponds to the
self-dual word $(X,\bar X)$, whose successive cyclic length-$n$
windows are $T^bX$.

\begin{lem}[{\cite[Theorem~15]{EP}}]\label{lem:seed-construction}
Let $X_0,\ldots,X_{r-1}\in\F_2^n$ belong to distinct full $T$-orbits.
Assume that
\begin{equation}\label{eq:seed-conditions}
 d_H(X_i,X_{i+1})=1\ (0\le i<r-1),\ d_H(X_{r-1},TX_0)=1.
\end{equation}
Then the words in the following array, read from left to right
in each row and from top to bottom, form a single-track Gray code
of length $n$ and period $P=2nr$:
\begin{equation}\label{eq:code-words}
\begin{array}{cccc}
 X_0         & X_1         & \cdots & X_{r-1}         \\
 TX_0        & TX_1        & \cdots & TX_{r-1}        \\
 T^2X_0      & T^2X_1      & \cdots & T^2X_{r-1}      \\
 \vdots      & \vdots      &        & \vdots          \\
 T^{2n-1}X_0 & T^{2n-1}X_1 & \cdots & T^{2n-1}X_{r-1}
\end{array}
\end{equation}
The codeword at index $br+i$ is $C_{br+i}=T^bX_i$
for $0\le b<2n$ and $0\le i<r$.
The head positions are $h_j=jr\pmod P$ for $0\le j<n$.
\end{lem}

\subsection{Directed multigraphs and transition systems}
\label{subsec:digraphs}

All graphs considered here are finite. We refer to \cite{BM} for
standard graph terminology. A \emph{directed multigraph}
$\mathcal D=(V,\mathcal E)$ has a vertex set $V$, an arc set
$\mathcal E$, and maps
\[
 \operatorname{tail},\ \operatorname{head}:\mathcal E\longrightarrow V.
\]
An arc is directed from its tail to its head. Distinct arcs may have the
same tail and head; these are \emph{parallel arcs}. An arc whose tail
equals its head is a \emph{loop}. A vertex is incident with an arc if it
is its tail or its head.

For $x\in V$, the incoming and outgoing \emph{ports} are the tagged copies
\begin{align*}
 I_x&=\{a^{\mathrm{in}}:a\in\mathcal E,\ 
                   \operatorname{head}(a)=x\},\\
 O_x&=\{a^{\mathrm{out}}:a\in\mathcal E,\ 
                   \operatorname{tail}(a)=x\}.
\end{align*}
The tags make these sets disjoint. A loop at $x$ contributes one port to
each set. The \emph{indegree} and \emph{outdegree} of $x$ are
$d^-(x)=|I_x|$ and $d^+(x)=|O_x|$, respectively, with parallel arcs
counted separately. The digraph is \emph{balanced} if these degrees are
equal at every vertex.

A \emph{directed walk} is a sequence of arcs $a_0,\ldots,a_{k-1}$
such that $\operatorname{head}(a_i)=\operatorname{tail}(a_{i+1})$
for $0\le i<k-1$. It is a \emph{trail} if its arcs are distinct.
A \emph{directed circuit} is a nonempty closed directed trail, so its
last arc ends at the tail of its first arc. Vertices may repeat in a
trail or a circuit. An \emph{Euler circuit} is a directed circuit
containing every arc exactly once.

\begin{defi}\label{def:transition-system}\normalfont
A \emph{transition system} on $\mathcal D$ is a family of bijections
$\tau_x:I_x\longrightarrow O_x$, one for each vertex $x$. Its successor
map $\pi:\mathcal E\longrightarrow\mathcal E$ is defined by
\[
 \pi(a)=b\ \Longleftrightarrow\ 
 \tau_{\operatorname{head}(a)}(a^{\mathrm{in}})=b^{\mathrm{out}}.
\]
\end{defi}

Each arc has one successor and one predecessor, so $\pi$ is a
permutation. Its cycles partition the arc set into directed circuits.
The transition system gives an Euler circuit precisely when $\pi$ has
one cycle. Prescribing the permitted pairs of incoming and outgoing
ports imposes local restrictions on this circuit decomposition.

\subsection{Bipartite graphs and matchings}\label{subsec:matchings}

A graph is \emph{bipartite} with parts $L$ and $R$ if its vertex set is
$L\sqcup R$ and every edge joins the two parts. A \emph{matching} is a
set of edges with no common endpoints. It is \emph{perfect} if it covers
every vertex. Thus a perfect matching in a bipartite graph specifies a
bijection from $L$ to $R$ using edges of the graph. For a multigraph,
degrees count parallel edges separately; its \emph{underlying simple
graph} retains one edge for each adjacent pair of distinct vertices.
A graph is \emph{$d$-regular} if every vertex has degree $d$.

For $U\subseteq L$, let $N(U)\subseteq R$ be its set of neighbours.
Hall's theorem \cite[Theorem~16.4]{BM} states that a matching covering
$L$ exists if and only if
\[
 |N(U)|\ge |U|\ \text{for every }U\subseteq L.
\]
When $|L|=|R|$, such a matching is perfect. The criterion applies also
to bipartite multigraphs. In particular, a $d$-regular bipartite
multigraph, $d\ge1$, has a perfect matching, and its edge set is a
disjoint union of $d$ perfect matchings. This decomposition will be used
at the zero vertex in Section~\ref{sec:transitions}.

\subsection{Perfect matchings in the hypercube}

The \emph{$d$-dimensional hypercube} $Q_d$ has vertex set $\F_2^d$,
with two words adjacent when they differ in exactly one coordinate.
It is a $d$-regular bipartite graph, with the even-weight and odd-weight
words as its two parts. Let $e_i$ denote the word with a one in coordinate
$i$ and zeros elsewhere. For each coordinate $i$, the edges
$\{x,x+e_i\}$ form a perfect matching, called a \emph{coordinate
matching}. An isomorphism between simple graphs is a bijection of their
vertex sets that preserves adjacency in both directions.

Fixing one coordinate to zero or one gives two subgraphs of $Q_d$,
each isomorphic to $Q_{d-1}$. The edges between them form the
corresponding coordinate matching.

A \emph{Hamilton cycle} visits every vertex exactly once before
returning to its initial vertex. Two edge-disjoint perfect matchings on
the same vertex set form a union of even cycles, with edges alternating
between the two matchings. We will use a matching whose union with a
given matching is a single Hamilton cycle.

Let $K(Q_d)$ denote the complete graph on $V(Q_d)$, whose edges are all
unordered pairs of distinct vertices. Fink's theorem applies to any
perfect matching of this complete graph; the added matching consists
of hypercube edges.

\begin{thm}[Fink~\cite{Fink}]\label{thm:fink}
Let $d\ge2$ and let $\mathcal P$ be a perfect matching of $K(Q_d)$.
There is a perfect matching $\mathcal M$ of $Q_d$ for which
$\mathcal P\cup\mathcal M$ is a Hamilton cycle on $V(Q_d)$.
\end{thm}

Algorithms for finding $\mathcal M$ are given in \cite{FinkAlg}.

\section{Complemented rotation orbits}\label{sec:orbits}

At length $2p$, each full $T$-orbit contributes $4p$ words to the
construction in Lemma~\ref{lem:seed-construction}. To attain period
$2^{2p}-4$, the chosen orbits must therefore cover all but four of the
binary words of length $2p$. We begin by identifying these orbits.

Throughout the remainder of the paper, $p$ is an odd prime, $n=2p$, and
$r=(4^{p-1}-1)/p$. Coordinates of $p$-words are indexed by $\Z/p\Z$.
Let $R$ denote left rotation, $(Ru)_i=u_{i+1}$, and let $e_i$ be the
$i$th unit vector. Put
\[
 A=\{u\in\F_2^p:\wt(u)\equiv0\pmod2\}.
\]
For $u\in A$, write $[u]=\{R^ju:j\in\Z/p\Z\}$. For a pair
$(u,v)\in A^2$, its diagonal rotation class is
\[
 [(u,v)]=\{(R^ju,R^jv):j\in\Z/p\Z\}.
\]

\subsection{A change of coordinates}

Grouping the coordinates in pairs $(x_i,x_{p+i})$ makes the action of
$T$ a cyclic shift of the $p$ pairs, except that the pair $(a,b)$
crossing the boundary becomes $(b,\bar a)$. We choose coordinates in
which the same rule applies at every position.

Let $\psi:\Z/4\Z\longrightarrow\F_2^2$ be the Gray map
\[
 \psi(0)=00,\ \psi(1)=01,\ 
 \psi(2)=11,\ \psi(3)=10.
\]
The Lee distance between two elements of $\Z/4\Z$ equals the
Hamming distance between their images under $\psi$~\cite[Theorem~1]{Hammons}.
For $x\in\F_2^{2p}$ and $0\le i<p$, define $q_i$ by
$\psi(q_i)=(x_i,x_{p+i})$. Choose $c\in\{1,3\}$ with
$cp\equiv1\pmod4$, and put
\begin{equation}\label{eq:coordinate-change}
 z_i=q_i-ci\pmod4,\ (u_i,v_i)=\psi(z_i).
\end{equation}
This defines a map $\Phi:x\mapsto(u,v)$. On pairs of $p$-words, write
\[
 R(u,v)=(Ru,Rv),\ S(u,v)=(v,\bar u).
\]

\begin{lem}\label{lem:conjugacy}
The map $\Phi$ is a Hamming isometry and
\begin{equation}\label{eq:conjugacy}
 \Phi T\Phi^{-1}=RS^c.
\end{equation}
Moreover, $\langle RS^c\rangle=\langle R,S\rangle$ has order $4p$.
\end{lem}

\begin{pf}
Adding a constant in $\Z/4\Z$ rotates the four vertices of the Gray
square. Thus \eqref{eq:coordinate-change}, together with the pairing of
coordinates, preserves Hamming distance.

In quaternary coordinates, $T$ sends
$(q_0,\ldots,q_{p-1})$ to $(q_1,\ldots,q_{p-1},q_0+1)$. For
$i<p-1$, the new $z_i$ is $z_{i+1}+c$. At the last coordinate it is
$q_0+1-c(p-1)=z_0+c$, since $cp\equiv1\pmod4$. Addition of one in
each Gray square acts as $S$, proving \eqref{eq:conjugacy}. Since $c$
is odd, $S^c$ has order four and generates $\langle S\rangle$.
The permutations $R$ and $S^c$ commute and have coprime orders $p$
and $4$. Hence $\langle RS^c\rangle=\langle R,S\rangle$ has order $4p$.
\end{pf}

\subsection{Orbit classes}

In the new coordinates, $R$ preserves the parity of each component,
whereas complementation reverses it because $p$ is odd. This allows
us to describe the intersection of each $\langle R,S\rangle$-orbit
with $A^2$ using rotation alone.

\begin{lem}\label{lem:orbit-count}
The $T$-orbits on $\F_2^{2p}$ are in bijection with the diagonal rotation
classes in $A^2$. The class $[(0,0)]$ corresponds to an orbit of size four.
Every other class has $p$ representatives in $A^2$ and corresponds to a
full $T$-orbit. Hence there are $r$ full orbits.
\end{lem}

\begin{pf}
Complementation reverses the parity of a $p$-word. The four iterates of
an even-even pair under $S$ have parity pairs
$00,01,11,10$. Thus each $S$-orbit contains exactly one pair in $A^2$.
Rotation preserves parity, so every $\langle R,S\rangle$-orbit meets
$A^2$ in one diagonal rotation class.

If $R^ju=u$ for $0<j<p$, then $u$ is constant. The only constant word
in $A$ is zero. Therefore a pair in $A^2\setminus\{(0,0)\}$ has
$p$ distinct rotations. Its orbit under $\langle R,S\rangle$ has size
$4p$. The orbit of $(0,0)$ consists of
$(0,0),(0,\one),(\one,\one),(\one,0)$. Since $|A|=2^{p-1}$,
the number of full orbits is $(|A|^2-1)/p=r$.
\end{pf}

The four words in the short $T$-orbit are obtained by taking $2p$
consecutive symbols from each of the four shifts of the periodic sequence
$00110011\cdots$. The code constructed below contains all remaining
binary words of length $2p$.

\section{Transition systems and circuit joining}\label{sec:transitions}

We next order the full orbits by constructing an Euler circuit in a
directed multigraph. The transitions of this circuit will determine the
one-coordinate changes between orbit representatives.

\subsection{The directed multigraph}

For a pair $(u,v)$, we take $[u]$ and $[v]$ as the endpoints of an
arc. The arc must also record the relative rotation of the two words,
since different diagonal classes can have the same endpoints.

Let $\mathcal D$ have vertex set $A/\langle R\rangle$ and arc set
\[
 \mathcal E=\{[(u,v)]:(u,v)\in A^2\setminus\{(0,0)\}\}.
\]
For an arc $a=[(u,v)]$, define its tail and head by
\[
 \operatorname{tail}(a)=[u],\ \operatorname{head}(a)=[v].
\]
Distinct diagonal classes give distinct arcs, including when their tails
and heads agree. A class with $[u]=[v]\ne[0]$ is a loop. The vertex
$[0]$ has no loop. Every two distinct vertices are joined by arcs in both
directions, and $|\mathcal E|=r$ by Lemma~\ref{lem:orbit-count}.

We use the incoming and outgoing port sets $I_x$ and $O_x$ from
Section~\ref{subsec:digraphs}.
When two arcs meet at $[v]$, their representatives can be rotated so
that the two middle words agree. We also need a condition on the
outer words to obtain a one-bit change between codewords.

Put $g_i=\one+e_i\in A$. Suppose that
$\operatorname{head}(a)=\operatorname{tail}(b)=x$. We permit the
transition from $a$ to $b$ if, for some $i\in\Z/p\Z$, representatives
can be chosen in the form
\begin{equation}\label{eq:transition}
 a=[(u,v)],\ b=[(v,w)],\ w=u+g_i.
\end{equation}
Let $H_x$ be the bipartite graph with parts
$I_x$ and $O_x$ in which $a^{\mathrm{in}}$ is adjacent to
$b^{\mathrm{out}}$ precisely when this transition is permitted. Thus
an edge of $H_x$ specifies a possible successor of an incoming arc.

The relation to Hamming adjacency is
\begin{equation}\label{eq:one-bit}
 S^{-1}(v,u+g_i)=(u+e_i,v),\ 
 d_H\bigl((u,v),S^{-1}(v,u+g_i)\bigr)=1.
\end{equation}
This identity will be used when the Euler circuit is converted into
binary codewords.

A transition system on $\mathcal D$ is \emph{permitted} if each
bijection $\tau_x$ pairs ports joined by an edge of $H_x$. Equivalently,
it is a choice of a perfect matching $M_x$ in $H_x$ for every vertex
$x$. We seek a permitted transition system whose successor permutation
has one cycle.

\subsection{Transitions at nonzero vertices}

At a nonzero vertex $[v]$, fixing the middle word $v$ determines a
unique representative for each incoming or outgoing port. The outer
words can then serve as labels for the ports. Complementing the
outgoing labels turns the transition rule into adjacency in a
hypercube.

\begin{lem}\label{lem:local-cube}
For every nonzero vertex $x=[v]$, the graph $H_x$ is isomorphic to
$Q_p$. Under this isomorphism, incoming ports have even-weight labels
and outgoing ports have odd-weight labels.
\end{lem}

\begin{pf}
Fix a representative $v\ne0$. Since $v$ has $p$ distinct rotations,
each incoming arc has a unique representative $(u,v)$, and each outgoing
arc has a unique representative $(v,w)$. In each case the other word
ranges over $A$.

Label the port belonging to $(u,v)$ by $u$, and label the port belonging
to $(v,w)$ by $\bar w$. These labels are respectively all even-weight
and all odd-weight words of length $p$. The relation $w=u+g_i$ is
equivalent to $\bar w=u+e_i$. Hence two ports are adjacent in $H_x$
if and only if their labels are adjacent in $Q_p$.
\end{pf}

In particular, the edges in any one coordinate direction of $Q_p$
give a local perfect matching at a nonzero vertex.

\subsection{A matching at the zero vertex}

At $[0]$, every rotation fixes the middle word, so this word does
not determine the relative alignment of the outer words. We shall
choose a matching containing a transition for which two alignments,
one rotation apart, are both permitted. This choice will be used to
close the sequence of generating words in Section~\ref{sec:construction}.

The incoming arcs at $[0]$ have the form $[(u,0)]$ and the outgoing
arcs have the form $[(0,w)]$, where $u,w\in A\setminus\{0\}$.
Their ports are therefore indexed on each side by the nonzero rotation
classes in $A$. Let
\[
 \sigma=[g_0],\ \beta=[e_0+e_1].
\]

\begin{lem}\label{lem:zero-matching}
There is a perfect matching at $[0]$ containing the transition
\begin{equation}\label{eq:distinguished-transition}
 [(g_0,0)]\longrightarrow[(0,e_0+e_1)].
\end{equation}
With the incoming representative $(g_0,0)$ fixed, this transition has
the two alignments
\begin{equation}\label{eq:two-alignments}
 w=e_0+e_1=g_0+g_1,\ 
 Rw=e_{p-1}+e_0=g_0+g_{p-1}.
\end{equation}
\end{lem}

\begin{pf}
Consider a bipartite multigraph $B$ whose parts are copies of the
nonzero rotation classes in $A$. For each left class $[u]$, fix one
representative $u$ and include an edge to $[u+g_i]$ for every $i$ such
that $u+g_i\ne0$. Replacing $u$ by a rotation permutes these edges,
so their multiplicities are independent of the chosen representative.
The underlying simple graph is $H_{[0]}$.

Every left vertex has degree $p$, except $\sigma$, which has degree
$p-1$. To compute the right degrees, let $m(U,W)$ be the number of
edges from a left class $U$ to a right class $W$. There are
$p\,m(U,W)$ pairs $(u,i)$ with $u\in U$ and $u+g_i\in W$.
The map $(u,i)\mapsto(u+g_i,i)$ is a bijection with the pairs counted
by $p\,m(W,U)$. Thus $m(U,W)=m(W,U)$, and the right degrees have the
same values.

Add one auxiliary edge between the two copies of $\sigma$. The
resulting graph is $p$-regular and bipartite, so its edges decompose
into $p$ perfect matchings by the result recalled in
Section~\ref{subsec:matchings}.
The edge from $\sigma$ to $\beta$ obtained from $g_0+g_1=e_0+e_1$
belongs to one of these matchings. This matching excludes the auxiliary
edge, since both edges use the left copy of $\sigma$. It is therefore
a perfect matching using only permitted transitions.

The identities in \eqref{eq:two-alignments} show that both choices
represent the same outgoing arc and give permitted transitions from
$(g_0,0)$. The two representatives are distinct. When $p=3$,
$\sigma=\beta$; the same argument applies with the parallel edges
retained in $B$.
\end{pf}

Fix such a matching at $[0]$ for the remainder of the construction.

\subsection{Joining the circuits}

The local matchings partition the arcs into circuits. To join the
circuits passing through a nonzero vertex $x$, we keep the trails
between successive visits to $x$ and choose new transitions between
their endpoints. The cube structure at $x$ allows us to use Fink's
theorem to make these trails into a single circuit.

\begin{lem}\label{lem:joining}
Let a permitted transition system on $\mathcal D$ be given, and let $x\ne[0]$.
The matching at $x$ can be replaced so that all circuits incident with
$x$ are joined into one circuit. Every other circuit is preserved.
\end{lem}

\begin{pf}
Delete the transitions at $x$. Each circuit incident with $x$ is cut
into directed trails. A trail starts at an outgoing port of $x$, follows
the corresponding arc and the retained transitions, and ends at the
next incoming port of $x$. These trails partition the arcs of the
affected circuits. Every port occurs as an endpoint of exactly one
trail. In particular, a loop at $x$ gives a trail consisting of one arc.

Pair the two endpoints of each trail. This gives a perfect matching
$\mathcal P$ on $I_x\sqcup O_x$. By Lemma~\ref{lem:local-cube}, these
ports are the vertices of $Q_p$, and each trail is directed from an
odd-weight label to an even-weight label. Apply
Theorem~\ref{thm:fink} to obtain a cube matching $\mathcal M$ such that
$\mathcal P\cup\mathcal M$ is a Hamilton cycle of $K(Q_p)$.

Orient the edges of $\mathcal M$ from incoming to outgoing ports.
The alternating cycle then traverses every edge of $\mathcal P$ from
an outgoing port to an incoming port. Replacing each such edge by its
directed trail gives one directed circuit containing all arcs of the
affected circuits. Its transitions at $x$ are the edges of
$\mathcal M$, so they are permitted. All transitions outside $x$
retain their previous assignments, which preserves every circuit
disjoint from $x$.
\end{pf}

Once two arcs belong to the same circuit, subsequent joins keep them
together. Every arc has a nonzero endpoint, and any two distinct
nonzero vertices share an arc. We can therefore carry out all the
joins at nonzero vertices and retain the prescribed transition at
$[0]$.

\begin{prop}\label{prop:euler}
The graph $\mathcal D$ has an Euler circuit consisting entirely of
permitted transitions and containing
\eqref{eq:distinguished-transition}.
\end{prop}

\begin{pf}
At each nonzero vertex choose a coordinate matching of the local cube,
and at $[0]$ use the fixed matching from Lemma~\ref{lem:zero-matching}.
Process the nonzero vertices one at a time, applying
Lemma~\ref{lem:joining}. Each operation replaces a collection of
circuits by a circuit having their union as its arc set. Hence, after a
vertex has been processed, all its incident arcs remain in one circuit
throughout the subsequent operations.

Any two distinct nonzero vertices share an arc of $\mathcal D$.
Therefore the circuits containing their incident arcs coincide after
all nonzero vertices have been processed. Every arc has a nonzero
endpoint, so all arcs belong to this one circuit. This reasoning also
covers the case of a single nonzero vertex. The matching at $[0]$ has
been retained, including \eqref{eq:distinguished-transition}.
\end{pf}

\section{Construction and examples}\label{sec:construction}

We now choose representatives along the Euler circuit and use them in
Lemma~\ref{lem:seed-construction}.

\subsection{The closing rotation}

The Euler circuit returns to its first arc. When we follow it with
aligned representatives, the final pair may be a rotation of the
initial pair. The two alignments at the prescribed transition let us
keep the arc order and choose representatives with a nonzero closing
rotation.

\begin{lem}\label{lem:closing-rotation}
Let $\alpha_0,\ldots,\alpha_{r-1}$ be an Euler circuit from
Proposition~\ref{prop:euler}, starting with
\[
 \alpha_0=[(g_0,0)],\ \alpha_1=[(0,e_0+e_1)].
\]
There are representatives $b_i$ of $\alpha_i$, $0\le i<r$, and a
representative $b_r$ of $\alpha_0$, such that every step
$b_i\longrightarrow b_{i+1}$ has the form \eqref{eq:transition} and
\begin{equation}\label{eq:closing-rotation}
 b_0=(g_0,0),\ b_r=R^\delta b_0
\ \text{for some }\delta\in(\Z/p\Z)\setminus\{0\}.
\end{equation}
\end{lem}

\begin{pf}
Begin with $b_0=(g_0,0)$ and $b_1=(0,w)$, where $w=e_0+e_1$.
Suppose that $b_i$ has been chosen. A permitted transition from
$\alpha_i$ to $\alpha_{i+1}$ has aligned representatives
$(u,v),(v,u+g_j)$. Since $(u,v)$ and $b_i$ represent the same arc,
a simultaneous rotation sends $(u,v)$ to $b_i$. Apply the same
rotation to the outgoing representative to obtain $b_{i+1}$.
Rotation permutes the vectors $g_j$, so the resulting step is permitted.
Continue through the closing transition to obtain
$b_r=R^\delta b_0$ for some $\delta\in\Z/p\Z$.

There is a second choice given by $b'_0=b_0$ and $b'_i=Rb_i$ for
$1\le i\le r$. Its first step uses the second alignment in
\eqref{eq:two-alignments}; each subsequent step is a rotation of the
corresponding original step. Its closing rotation is $\delta+1$.
At least one of $\delta$ and $\delta+1$ is nonzero. Choose the
representatives giving a nonzero closing rotation.
\end{pf}

To obtain the generating words, we first apply $S^{-i}$ to the
representative at step $i$. By \eqref{eq:one-bit}, successive pairs
then differ in one bit. Since $p$ is prime, the nonzero closing
exponent $\delta$ is invertible modulo $p$. A relabeling of coordinates
therefore changes $R^\delta$ to $R$ and gives the closing step to
$TX_0$ in the original binary coordinates.

\begin{prop}\label{prop:generating-words}
There are words $X_0,\ldots,X_{r-1}$, one from each full $T$-orbit,
which satisfy \eqref{eq:seed-conditions}.
\end{prop}

\begin{pf}
Take the representatives in Lemma~\ref{lem:closing-rotation} and set
\[
 y_i=S^{-i}b_i\ (0\le i\le r).
\]
Since $S$ is an isometry, \eqref{eq:one-bit} implies
$d_H(y_i,y_{i+1})=1$. The closing relation is
\begin{equation}\label{eq:transformed-closing}
 y_r=R^\delta S^{-r}y_0.
\end{equation}
From $pr=4^{p-1}-1$ and $cp\equiv1\pmod4$, we obtain
$-r\equiv c\pmod4$.

Let $a=\delta^{-1}$ in $\Z/p\Z$, and define a coordinate
permutation on $p$-words by
\[
 (D_a u)_i=u_{a^{-1}i}.
\]
Apply $D_a$ to both components of a pair. Direct calculation gives
$D_aR=R^aD_a$ and $D_aS=SD_a$. Thus
\begin{equation}\label{eq:unit-closing}
 D_ay_r=RS^cD_ay_0=(\Phi T\Phi^{-1})D_ay_0.
\end{equation}
Write $G=\langle R,S\rangle$. Since $a$ is nonzero modulo $p$,
$\langle R^a,S\rangle=G$. The relations above therefore give
\[
 D_a(Gy)=G(D_ay)
\]
for every pair $y$. Hence $D_a$ permutes the $G$-orbits and preserves
their sizes. The pairs $b_0,\ldots,b_{r-1}$ represent all full
$G$-orbits, and applying $S^{-i}$ to $b_i$ keeps it in the same orbit.
Consequently the words
\begin{equation}\label{eq:generating-words}
 X_i=\Phi^{-1}D_ay_i\ (0\le i<r)
\end{equation}
represent all full $T$-orbits. Both $D_a$ and $\Phi$ preserve Hamming
distance, so successive $X_i$ differ in one coordinate. By
\eqref{eq:unit-closing}, the last word is also at distance one from $TX_0$.
\end{pf}

\Needspace{14\baselineskip}
\subsection{The main theorem}\label{subsec:main-theorem}

The generating words from Proposition~\ref{prop:generating-words} give
codes attaining the period bound from Proposition~\ref{prop:bound}.

\begin{thm}\label{thm:main}
Let $p$ be an odd prime and let $r=(4^{p-1}-1)/p$. There exists a binary
single-track Gray code of length $2p$ and period
\[
 P=4pr=2^{2p}-4
\]
with head positions $h_j=jr\pmod P$, $0\le j<2p$. This period is the
largest possible for a binary single-track Gray code of length $2p$.
\end{thm}

\begin{pf}
Apply Lemma~\ref{lem:seed-construction} to the words from
Proposition~\ref{prop:generating-words}. The resulting code has length
$2p$, period $4pr=2^{2p}-4$, and head positions $h_j=jr\pmod P$.
Proposition~\ref{prop:bound} proves its optimality.
\end{pf}

\subsection{Construction algorithm}

The construction in Theorem~\ref{thm:main} can be implemented as follows.
Choices of representatives and matchings can be made in lexicographic order.

\begin{enumerate}
\item Enumerate the even-weight words in $A$ and their rotation classes
(One can generate the lexicographically least representative
of each length-$p$ binary necklace using the algorithm analyzed
in~\cite{RuskeySavageWang} and retain the representatives of even weight).
Construct the arc set $\mathcal E$. For arcs with nonzero first
component, take the least representative $u$ of each nonzero class and
all pairs $(u,v)$ with $v\in A$. For arcs with first component zero,
take $(0,v)$ with $v$ the least representative of each nonzero class.

\item Construct a perfect matching at $[0]$ containing
\eqref{eq:distinguished-transition}. This can be done by fixing the
edge $\sigma\beta$, deleting its endpoints, and applying a bipartite
matching algorithm~\cite{HopcroftKarp} to the remaining permitted transitions.

\item At every nonzero vertex, choose a coordinate matching of the
local cube. Process these vertices successively. At the current vertex,
form the endpoint matching of the exterior trails and find its cube
completion as in Theorem~\ref{thm:fink}. Use this completion as the
new local matching. A vertex whose incident arcs already belong to
one circuit can be skipped.

\item Read the resulting Euler circuit starting with
\eqref{eq:distinguished-transition}. Choose the representatives
$b_0,\ldots,b_r$ as in Lemma~\ref{lem:closing-rotation}. Determine
$\delta$, choose the alignment giving $\delta\ne0$, and calculate
the generating words from \eqref{eq:generating-words}.

\item Store $X_0,\ldots,X_{r-1}$. For a row index $k$, reduce $k$
modulo $P$, write $k=br+i$ with $0\le i<r$, and output $T^bX_i$.
Its zeroth coordinate is the corresponding symbol of the generating
track.
\end{enumerate}

The matching completion in Step~3 can be implemented recursively
\cite{Fink,FinkAlg}. Split a cube along a coordinate crossed by an edge
of the given matching. In one half, pair the exposed endpoints
temporarily and recursively complete the resulting perfect matching.
Deleting the temporary pairs leaves trails whose endpoints, together
with the crossing edges, induce a matching on the exposed vertices of
the second half. Complete the matching in that half and expand the
induced pairs. The recursion ends in dimension two, where the two
coordinate matchings can be tested directly.

The stored list has $r$ binary words of length $2p$, while the full code
has $4pr$ words. Thus the generating-word representation reduces the
number of stored bits by a factor of $4p$.

\subsection{Examples}

To check adjacency in the array \eqref{eq:code-words}, it suffices to
check the steps within the first row and the step from its last word
to the first word of the next row. Applying $T$ gives all the other
steps. Both checks can be seen directly in the smallest case.

\begin{ex}\label{ex:length-six}\normalfont
Let $p=3$. Then $r=5$, and a choice of generating words is
\[
 \texttt{001000},\ \texttt{001001},\ \texttt{000001},\ 
 \texttt{000101},\ \texttt{010101}.
\]
Successive words differ in one coordinate. The last word is also at
distance one from $T(\texttt{001000})=\texttt{010001}$. These five
words belong to distinct full $T$-orbits, each of size twelve.
Formula~\eqref{eq:code-words} therefore gives a code of length six
and period $60$, with head positions $0,5,10,15,20,25$.
\end{ex}

Table~\ref{tab:parameters} gives the parameters of the construction for
the first five odd primes. The values follow from
$r=(4^{p-1}-1)/p$ and $P=4pr$. For $p=17$, the same construction gives
period $17{,}179{,}869{,}180$.

\begin{table}[htbp]
\centering
\caption{Parameters of the codes in Theorem~\ref{thm:main}.}
\label{tab:parameters}
\begin{tabular}{rrrr}
\toprule
$p$ & Length $2p$ & Generating words $r$ & Period $P$\\
\midrule
3 & 6 & 5 & 60\\
5 & 10 & 51 & 1,020\\
7 & 14 & 585 & 16,380\\
11 & 22 & 95,325 & 4,194,300\\
13 & 26 & 1,290,555 & 67,108,860\\
\bottomrule
\end{tabular}
\end{table}

A generated code can be checked by verifying that its words are
distinct, that consecutive words have Hamming distance one, including
the last and first, and that each column is the prescribed shift of
column zero. For the generating words, it suffices to check that their $T$-orbits
are full and pairwise distinct and that \eqref{eq:seed-conditions} holds.

\FloatBarrier
\section{Conclusion and open problem}\label{sec:conclusion}

In this paper, we have constructed a binary STGC of length
$2p$ from full-order self-dual necklaces of length $4p$ for every odd prime $p$, as suggested
by Schwartz and Etzion~\cite{SE}. Its period $P=2^{2p}-4$ attains the
upper bound in Proposition~\ref{prop:bound}. This proves the conjecture
stated in the introduction and gives an infinite family of optimal
binary STGCs. We use Fink's matching theorem~\cite{Fink} at the
nonzero vertices of a directed multigraph to obtain an Euler circuit
that orders the full $T$-orbits. The choice of alignment at the zero
vertex ensures that, after relabeling the coordinates, the last
generating word differs in one bit from the $T$-shift of the first.
The resulting code is specified by $r=(4^{p-1}-1)/p$ generating words
and has head positions $h_j=jr\pmod P$ for $0\le j<2p$.

The period formula $2^{2m}-4$ does not extend to all odd composite
$m$. By Proposition~\ref{prop:bound}, the period of a binary STGC of
length $2m$ must be divisible by $4m$. Thus $2^{2m}-4=4(4^{m-1}-1)$
can be attained only if $4^{m-1}\equiv 1\pmod m$. This congruence
holds for every odd prime $m$ by Fermat's little theorem, but it
already fails for $m=9$, since $4^8\equiv 7\pmod 9$. When $m$ is composite, the construction faces a further difficulty. A nontrivial
rotation can fix a nonzero even-weight word of length $m$, so fixing
the middle word may leave several choices of representative for an
incident arc. Moreover, normalizing the closing rotation requires its
exponent to be relatively prime to $m$. It remains an open problem
to determine the maximum period of a binary STGC of length $2m$ for
arbitrary odd composite $m$ and to construct codes attaining it.\newline

\noindent
{\bf Acknowledgement:} The authors wish to express their sincere gratitude to Professor Etzion for suggesting the research problem and for his valuable and insightful guidance during his visit to Hefei in November 2025.

\noindent
{\bf Declaration of AI use:} The authors used ChatGPT (OpenAI) for proofs and language editing. All the AI-assisted content was reviewed, verified, and revised by the authors, who take full responsibility for the results, the arguments, and the final form of the article.

\end{document}